\documentclass[journal]{IEEEtran}

\usepackage{balance}
\usepackage[normalem]{ulem} 

\usepackage{listings} 
\usepackage{amsmath}
\usepackage{subcaption}
\usepackage{graphicx}
\graphicspath{{./images/}}
\usepackage{color}

\usepackage{algorithm} 

\usepackage{substitutefont}
\usepackage{ifpdf}
\usepackage[T1]{fontenc}
\usepackage[utf8]{inputenc}

\usepackage{algcompatible}
\usepackage{algpseudocode}
\usepackage[shortlabels]{enumitem}
\usepackage{xspace}
\usepackage{hyperref}

\usepackage[style=ieee]{biblatex}
\usepackage{booktabs}
\usepackage{multirow}
\usepackage{setspace} 
\usepackage{pifont}
\usepackage{outlines}
\usepackage{adjustbox}
\usepackage{multirow}
\usepackage{tabularx}
\usepackage{makecell}
\usepackage{rotating}
\usepackage{enumitem}
\usepackage{tikz} 
\usetikzlibrary{tikzmark}
\usetikzmarklibrary{listings}
\usepackage{xcolor}
\usepackage{msc}
\usepackage[htt]{hyphenat}
\usepackage[zerostyle=b,scaled=1]{newtxtt}

\usepackage{framed}
\usepackage{tcolorbox}
\usepackage{newverbs}
\usepackage{soul}
\usepackage{cleveref}
\crefname{section}{§}{§§}

\newcounter{tmkcount}
\tikzset{
  use tikzmark/.style={
    remember picture,
    overlay,
    execute at end picture={
      \stepcounter{tmkcount}
    },
  },
  tikzmark suffix={-\thetmkcount}
}

\newcommand*{\hdr}[1]{\noindent\textbf{#1}\xspace}

\newcommand*{\thename}[0]{\texttt{PIM-Enclave}\xspace}

\newcommand*{\hj}[1]{#1}

\lstnewenvironment{code}[1]{\lstset{ language=#1,
    directivestyle={\color{black}}, emph={},
    keywordstyle=\color{blue}, basicstyle=\footnotesize\ttfamily,
    stringstyle={\color{black}}, frame=None, showstringspaces=false,
    xleftmargin=0.3cm, framexleftmargin=0.3cm,
    emphstyle={\color{blue}},
    keywordstyle=\color{black}\bfseries, 
    commentstyle={\color{blue}},
    morekeywords={LOTR_SECRET,PRIVUSER_SECRET,__privcall2,lcall,lret,asm,privcall,
      PRIV_SECRET,syscall,PRIVCALL_DEFINE,SYSCALL_DEFINE},
    escapechar={\^}, }}{}

\begin{document}
%
\title{\Large \bf PIM-Enclave: Bringing Confidential Computation Inside Memory}
%
%
%
\author{Kha Dinh Duy,
        Hojoon Lee*\thanks{*Corresponding author} \\
        \{khadinh,hojoon.lee\}@skku.edu \\
        Department of Computer Science and Engineering, \\ Sungkyunkwan University
        
        }

\maketitle

\begin{abstract}
Demand for data-intensive workloads and confidential computing are the prominent research directions shaping the future of cloud computing. Computer architectures are evolving to accommodate the computing of large data better. Protecting the computation of sensitive data is also an imperative yet challenging objective; processor-supported secure enclaves serve as the key element in confidential computing in the cloud. However, side-channel attacks are threatening their security boundaries. The current processor architectures consume a considerable portion of its cycles in moving data. Near data computation is a promising approach that minimizes redundant data movement by placing computation inside storage. In this paper, we present a novel design for Processing-In-Memory (PIM) as a data-intensive workload accelerator for confidential computing. Based on our observation that moving computation closer to memory can achieve efficiency of computation and confidentiality of the processed information simultaneously, we study the advantages of confidential computing \emph{inside} memory. We then explain our security model and programming model developed for PIM-based computation offloading. We construct our findings into a software-hardware co-design, which we call \thename. Our design illustrates the advantages of PIM-based confidential computing acceleration. Our evaluation shows \thename can provide a side-channel resistant secure computation offloading and run data-intensive applications with negligible performance overhead compared to baseline PIM model.
\end{abstract}

\begin{IEEEkeywords}
processor-in-memory, confidential computing
\end{IEEEkeywords}

%
\IEEEpeerreviewmaketitle

%
%
%
%

\section{Introduction}

\IEEEPARstart{T}{}oday's cloud computing is facing two urgent challenges: improving efficiency in data-centric computation and providing confidentiality of sensitive data computation. Data-intensive workloads such as large data analytics and training of neural networks have become the most common use cases of the cloud. Recent advancements and future research directions in processor architectures, accelerators, and memory technology are also centered around this trend. Ensuring the confidentiality of the computation in the cloud is another agenda that is shaping the future of cloud computing. Many cloud service providers are already providing confidential computing services that adapt security extensions in modern processor architectures~\cite{google-confidential-computing,azure,nitro}.

Many previous works have proposed methods for securing data computation in the cloud based on commonly available processor-supported \emph{secure enclaves}~\cite{vc3,occulemency,opaque,oblivious-ml}. Secure enclaves such as Intel SGX~\cite{sgx} allow building of a strong security model in which the data and its computation are protected from possibly malicious cloud service providers and co-tenants. The code and data inside an enclave are only visible when a thread is in enclave mode and protected from any other software, including the OS kernel. Also, enclave-protected contents are encrypted as it leaves the CPU package to be stored in memory. Applications can be modified or developed to protect their sensitive code and data by placing them inside enclaves. For instance, VC3~\cite{vc3} retrofits the MapReduce framework to protect processed data, and Occlumency~\cite{occulemency} proposes SGX-protected deep learning inference.

Unfortunately, recent works have revealed that upholding the strong security model of secure enclaves amid surrounding threats proved to the much more complicated than initially considered. Researchers have shown that secure enclaves can be vulnerable to microarchitectural side-channel attacks~\cite{sgx-grand-exposure,sgx-cache-attack}, or controlled side-channel attacks~\cite{sgx-page-table,sgx-step,plundervolt} launched by system software that enclaves depend on for accessing system resources (e.g., Intel SGX model). Studies showed that such attacks are sufficient to compromise the confidentiality of the computation encapsulated by secure enclaves. In response, a number of works proposed defenses against the attacks. Oblivious RAM (ORAM)-based approaches for SGX have been explored by many works~\cite{obfuscuro, zerotrace,raccoon}. The approach transforms the protected program such that its memory access pattern does not leak any information about the program. However, applying ORAM to a program makes its memory access time and throughout several magnitudes slower than native. Trustore~\cite{trustore} proposes a PCI-based secure storage design as an alternative to the ORAM-based approaches, pointing out their impractically high performance overhead. Besides, secure enclaves in the cloud also have limited memory (e.g., 128MB in case of SGX). This is a critical bottleneck for data-intensive applications in terms of program performance and development efforts; developers have to pay a close attention to the limitation and write their programs in a way that data is loaded into enclave in smaller chunks~\cite{occulemency}.

Another struggle in the cloud is to overcome the data movement bottleneck, which is often referred to as the \emph{memory wall problem}. The size of the processed data sets is ever-increasing, and computer architectures and the cloud are evolving to become more efficient in computing over large data sets. Today's computer architectures spend more than half of their cycles on moving the data. The near data processing approach seeks to solve the problem by enabling computation inside storage devices to minimize data movement~\cite{ndp-framework,nest,medal,near-data-uncore, computing-with-near-data,beyond-the-wall}. \emph{Processor-In-Memory (PIM)} is one of the most prominent directions in the trend~\cite{enabling-practical-pim,lazypim,ndp-framework,in-memory-parallel}. The idea is to bring computation inside memory such that the computation enjoys very low-latency access to data residing in memory, and at the same time, minimize unnecessary data movement through the system bus. The approach is actively being explored by the industry~\cite{samsung-pim,upmem} and in academia~\cite{tesseract,pei,smcsim,enabling-practical-pim,lazypim}. PIM-accelerated computation of large data has shown its potential through many works~\cite{impica,nda,prime, chameleon, mr-ndc, tesseract,pei, graphpim, ambit}. Notably, a few works~\cite{invisimem,obfusmem,hega,sdimm} have explored the use of PIM for security. For instance, \cite{invisimem,obfusmem} proposed PIM-assisted address side-channel mitigation on the system bus.

In this work, we present a novel PIM-based in-memory enclave model based on our observation that PIM can simultaneously achieving confidentiality and efficiency in large data computation in the cloud. Moving data to computation (i.e., the main processor) creates a constraint on the memory channel and may create side-channels along with the way. We argue that bringing computation to memory instead of the opposite significantly reduces the attack surface against confidential computing. However, security requirements and applications of an enclave-inside-memory are underexplored. We explain our study on the security model and programming model for enclave functionality in PIM, and generalize them into a software-hardware co-design called \thename. Then, we show the merits of secure acceleration provided by PIM in two aspects. First, \thename can enclavize a memory vault to facilitate data protection inside memory and offload the data-intensive workload based on our security model. Second, \thename can hide the memory access pattern of sensitive workloads by computing them in memory, eliminating address side-channels in the memory bus and main processor caches.

We summarize the contributions of this work as the following:

\begin{itemize}
\setlength\itemsep{1pt}

\item This is the first work that explores the potentials of PIM as a secure computation accelerator and discusses its inherent advantages.

\item We present a novel design for confidential computing using PIM by studying the design space as well as a thorough security requirement analysis.

\item We propose a set of non-intrusive yet imperative modifications to the PIM architecture for ensuring confidentiality and integrity of computation inside memory.

\item We evaluate our design through a confidential computation of real-world data-intensive workload on our architecture to show that our proposal for PIM-based confidential computing achieves both efficiency and security.

\end{itemize}


\section{Background and Assumptions}

\subsection{Processing-In-Memory}
PIM is a general direction towards embed processing capability inside memories to overcome the memory bandwidth bottleneck as today's workload is becoming more and more data-intensive. PIM architecture and programming models have not been standardized. We briefly explain the existing works on PIM from both the industry and academia. Then, we will explain the baseline PIM model, a generalized PIM architecture based on our observations, on which we build our architecture \thename.

\hdr{PIM hardware architectures.} The progress made in 3D-stacked memory technology enabled embedding of logics or even processors into memory.  Two of the most prominent example of 3D-stacked memory are \emph{Hybrid memory cube (HMC)}~\cite{hmc}, and \emph{High Bandwidth Memory (HBM)} \cite{amd-hbm,samsung-pim}. Stacked memory architectures vertically stack DRAM layers on top of each other and connect the vertical partitions of memory using high-bandwidth through-silicon vias (TSVs). A typical 3D-stacked memory configuration can employ thousands of TSVs \cite{primer-on-pim}, which makes its internal memory bandwidth far exceed that of traditional memory systems. At the bottom of the memory stacks, there is a \emph{logic layer} that can host hardware logic that can interact with both the host processor and the DRAM memory. A non-3D stacked PIM architecture also exists; UPMEM \cite{upmem} introduced its PIM architecture. At a high level, the UPMEM architecture couples each DRAM bank with a low-power RISC-like processor called the data processing unit (DPU) to perform general-purpose computation and have direct access to data in the bank.

\hdr{Accelerating computation with PIM.} 
The logic inside memory can be special-purpose hardware logic~\cite{invisimem, ambit,prime} or a general-purpose processor~\cite{impica, upmem, samsung-pim} that runs software often called the \emph{PIM kernels} analogous to the kernels loaded onto GPUs. PIM logics or kernels have a common purpose: perform operations that require frequent memory access such that unnecessary data movement is minimized.

\subsection{Baseline PIM Overview}
\label{subsec:pim-baseline}
Our work explores a design space for achieving confidential computing inside PIM architecture. Hence, it is necessary that we generalize the \emph{baseline} PIM architecture so that we can analyze the security requirements and construct a confidential computing architecture. There is a plethora of designs and implementations of PIM proposed in previous works in both the literature~\cite{pei,smcsim,tesseract,samsung-pim,computedram,mr-ndc,impica,near-data-acceleration,top-pim,neural-network-training,gradpim,graphpim} and industry~\cite{samsung-pim,hmc,amd-hbm,upmem}. The accumulating research has shaped a certain consensus on the design patterns and programming models, while many design choices are still open-ended research questions.  We outline the key design components common to PIM systems and compose them into a generalized PIM architecture based on such commonalities. 

\hdr{Baseline hardware configuration.}
We assume that a general-purpose low-power core is integrated into where memory is located. For 3D-stacked memory, the processing core is placed in the logic layer of each memory bank. For DRAM-based PIM such as~\cite{upmem}, the logic is coupled at the level of DRAM banks. For brevity, we will call the memory bank (in DRAM-based PIM) and the memory vault (in 3D-stacked PIM) generally as a \emph{memory bank}. We refer to the processor core inside memory as the \emph{PIM core}. As to the processing capacity of these processors, they often lack processor caches~\cite{upmem,smcsim,top-pim,neural-network-training}, and packaged with a low-latency on-chip memory, or \emph{PIM local memory}~\cite{smcsim,upmem,cho2020accelerating}. Therefore, there are two types of memories inside a PIM-enabled memory bank: the bank memory (in GBs, e.g., 64GB) and a relatively smaller PIM on-chip memory (a few MBs). The PIM core would fetch a batch size of data from the memory bank to its local memory via DMA to perform computation. However, each PIM core cannot access other memory banks but the one that it resides and have no means to communicate with other cores unless assisted by the host program. Lastly, PIM does not have permanent storage, such as a hard drive. 

\hdr{Communication with the host. }
Most PIM systems communicate with the host system via memory-mapped I/O and register-mapped I/O. Commonly, there is a \emph{command interface} at PIM-side that receives commands and allows the host to control PIM core execution. The control interface can be easily implemented in 3D-stacked memory such as HMC and HBM, as they can employ a handling logic inside the logic layer that can handle special commands. However, implementing such an interface on DRAM is proven complicated due to the constraints in the pin number of the DDR interface. The UPMEM architecture accomplishes the control interface by reserving a set of \emph{bank group} and \emph{bank address} of DDR4 for this interface; any accesses targeting this range of address is directly forwarded to the registers and memory of the PIM cores~\cite{upmem-openpower}.

\hdr{Programming model. }
In order to utilize PIM, the host would load PIM local memory with a program, often referred to as a \emph{PIM kernel}. The PIM kernel is a bare-metal program and lacks an operating system. The host and PIM both have the bank memory mapped into their address spaces. Data and parameters for commands are shared using the shared physically contiguous memory between host and PIM. This way, the PIM core can translate the host-provided pointer to the PIM's physical address view on the bank memory with an offset. In addition, we do not assume architectural support for cache coherency between the PIM core and host processor. Such a cache coherence scheme often requires PIM-aware changes to the host processor architectures. However, we do not involve concurrent computation on the same data between the PIM core and host processor in our current design. Typically, a host program running on the host processor is employed to control the execution of PIM cores. This program distributes data to the PIM-enabled memory banks, then sends commands to the control interface along with the parameters to command PIM to start execution. Then, the host program can either wait for PIM to finish by polling on the status register, or asynchronously work on other tasks. This programming model is similar to that of using GPUs and is also adapted by many previous works~\cite{upmem,tesseract,impica,neural-network-training}.

\subsection{Confidential Computing and Side-channel Attacks}

\hdr{Secure enclaves and the cloud. } Modern processor architectures provide support for a secure enclave for isolated execution of sensitive code and data. Intel SGX~\cite{sgx} is the most commonly available secure enclave in the cloud. SGX provides protected memory pages called \emph{Enclave Page Cache (EPC)} in which the sensitive subset of a program and processed data can be protected. A remote user can verify the integrity of an enclave through \emph{remote attestation}, a procedure that allows an enclave to \emph{attest} its code and data with a measurement, signed by a secure hardware cryptographic key. Previous works have leveraged SGX for secure data-intensive computation in the cloud to protect the workload within the possibly untrusted cloud~\cite{vc3,enclavedb,occulemency}. However, SGX has a limited EPC capacity of 128MB, and large data computation has to be broken down into smaller batches~\cite{occulemency}. Alternatively, a previous work~\cite{trustore} proposed using a PCIe-based storage device to extend the limited memory size of the secure enclave. 

\hdr{Side-channel attacks on enclaves and mitigation. } Researchers have shown that secure enclaves can fall vulnerable to side-channel attacks. The cache side-channels are shown to be exploited against SGX secure enclaves to undermine the confidentiality of the computed data~\cite{leaky-cauldron,sgx-cache-attack,sgx-grand-exposure}. Also, controlled side-channels exist due to the SGX model still interacting with the untrusted kernel for system resources~\cite{sgx-page-table,sgx-step}. Trustore~\cite{trustore} uses a narrow single-channel to interface a PCIe storage and SGX enclave to mitigate side-channels in processor caches and address side-channel on the bus. ORAM adaptation inside enclave~\cite{zerotrace} is another approach for memory side channels. However, due to its high memory performance overhead, it is inapt for data-intensive computing. Our work conducts a precursory study that explores a design space that combines efficient large-data computation (PIM) and confidential computation.

\section{Threat Model and Assumptions}
\label{sect:threat-model}

We assume a strong threat model in which the adversary controls system software in all privileged layers and has physical access to the system. We only trust hardware-protected enclaves such as our \thename and CPU-based enclaves. The primary objective of the adversaries is to extract sensitive information protected by secure enclaves through powerful attacks that they can launch with kernel privilege and physical access. 
Side-channel attacks on the host CPU that take advantage of microarchitectural characteristics are out of scope of our protection. However, the host enclave can offload sensitive operations to \thename to benefit from side-channel-free execution.
Moreover, we also deem denial of service attacks out of scope since they do not adhere to adversary objectives. Our threat model is in line with much previous work on secure enclaves and related research\cite{trustore,obfusmem,invisimem}.

\hdr{Malicious system software. } 
We place the system software outside our \emph{trusted computing base} (TCB), the same as other trusted execution environments.   
As the privileged software manages the memory mappings, a malicious system software could map a region of physical of memory into its own memory space or to a malicious process memory space. This also allows the attacker to passively or actively tap into every IO communications to DMA buffers of MMIO from the host with the PIM device. 

\hdr{Microarchitectural and controlled side-channels. } As demonstrated in numerous research works~\cite{sgx-cache-attack, sgx-grand-exposure,sgx-page-table,sgx-step}, processor-based enclaves suffer from information leakage through various side-channels. We consider side-channels caused by the host CPU out of the scope of this work, as the mitigation methods have been discussed in several works~\cite{raccoon,catalyst,new-cache-design-side-channel}. However, our model protects code and data accesses offloaded to \thename, which are side-channel-free by design.

\hdr{Physical attacks. }
The adversaries are assumed to have physical access to the hardware, allowing them to tap into the system bus. This leads to several kinds of physical attacks on the MMIO interface of devices and the memory system. First, the attacker might corrupt the packet by tampering with the payload. He might also replay or delay a previously valid packet on the bus. The attacker could perform access pattern analysis to infer sensitive information from the memory accesses or the MMIO packets \cite{access-pattern-disclosure,membuster}.  We also consider DMA attacks \cite{write-access-pattern}, where an attacker take snapshots of the memory content at a high frequency, and \emph{cold boot} attacks, where an attacker exploits the memory retention behavior DRAM to extract the content of memory directly. However, we assume that the tightly integrated PIM-enabled memory bank which includes the PIM core, PIM local memory and the access control logic are secure from physical attacks. Also, the connections inside the PIM hardware are often connected with highly integrated connections (e.g., Silicon Vias (TSV)), so we assume that using hardware probes to snoop data contents from PIM is infeasible.


\section{\thename Design}
We explain our design for a secure computation-in-memory model for the confidential acceleration of data-intensive workloads. We first explain the advantages of placing confidential computing into PIM in terms of security and performance (\cref{subsec:advantages}). Then, we describe an overview of \thename and how a user interacts with it (\cref{subsec:overview}).
We carefully analyze the attack vectors on PIM and devise a set of security requirements, which we use as the basis of our design (\cref{subsec:requirement}). We describe our support for attestation and how a secure session is established (\cref{subsec:attestation}). We then explain
the memory protection model, and how we achieve the confidentiality guarantees of data (\cref{subsec:memory}).
 Lastly, we also took considerations for performance with the introduction of the AES-capable DMA engine; our design should not hinder PIM's performance, which would defeat the purpose of our approach (\cref{subsec:aes-dma}).


\subsection{Motivations for PIM-based Confidential Computing} 
\label{subsec:advantages}
We briefly outline the potentials of a PIM-based enclave to convey our motivation for this work. The main advantages of \thename can be explained with respect to acceleration of large data computation and resilience against side-channels.

\hdr{Acceleration of data-intensive workloads.} 
Our work proposes a PIM model that offers a secure acceleration for data-intensive confidential computing, based on the findings of recent works on the advantages of PIM in data-heavy computations~\cite{scalable-pim-acc,impica,ambit, prime,chameleon,nda,biscuit}. Confidential computing inside memory solves problems that the current confidential computing in the cloud faces. Turning memory into an active entity in secure data storage and processing can solve the problem of limited memory space in cloud enclaves. This is a substantial problem that impedes the widespread adoption of confidential computing for large data. Existing works have proposed DNN model-based inference frameworks that are aware of and optimized for the SGX's memory limitations~\cite{occulemency,kim2020vessels}. A PCI-based secure storage device that complements the SGX's limitation was also proposed~\cite{trustore}. Our work seeks to explore a design in which memory cooperates with an in-processor enclave to store and process data. Moreover, many PIM designs promote dividing the data into multiple partitions, each processed by a different PIM processor~\cite{upmem,tesseract}. This also minimizes the attack surface, preserving the confidentiality of the rest of the data even when one partition got compromised.

\hdr{Mitigation of side-channels by design.} PIM can be naturally resistant to many side-channel attacks stem from \emph{resource sharing} that have been an existential threat to many processor enclaves~\cite{sgx-cache-attack,leaky-cauldron,sgx-page-table}. Performing computation where the data resides eliminates side-channels that occur while the data moves in route on the memory bus or stored in shared storage such as the processor caches. For instance, eliminating \emph{runtime} data movement from memory to processor inherently remove the address side-channels that arise on the memory bus~\cite{drama,membuster} that many previous ORAM-based approaches sought to mitigate~\cite{invisimem,obfusmem,zerotrace, freecursive-oram,trustore,sdimm}. Those approaches are effective but impractical for large-data computation due to their performance overhead. 





\begin{figure}[t]
%
\newcommand*\Reactivatenumber{%
  \lst@AddToHook{OnNewLine}{%
   \let\thelstnumber\origthelstnumber%
   \advance\c@lstnumber\@ne\relax}%
}

\newsavebox{\codeboxhost}
\newsavebox{\codeboxpim}

\begin{lrbox}{\codeboxhost}
\footnotesize
\begin{lstlisting}[language=C,name=codeL,escapechar=!,escapeinside=||,numbers=left]
// Executed inside host enclave 
PIMEnc* pe = PIMEncInit(PE_INDEX, 
                        PE_MODULE);
attest_device(pe);
establish_session(pe);
pe.load_binaries("pim.elf.enc");

// Allocate objects in memory bank
void *A_ptr = pe.alloc(sz);
pe.load(A_ptr, data_source, sz);

// Protect the memory region
pe.protect(objects,sz);

PIMParams params;
params[0] = A_ptr;

pe.offload_params(params);
pe.execute();
\end{lstlisting}
\end{lrbox}

\begin{lrbox}{\codeboxhost}
\scriptsize
\begin{lstlisting}[language=C,name=codeL,escapechar=!,escapeinside=||,numbers=left]
void *out_ptr = pe.get_output();
\end{lstlisting}
\end{lrbox}
\begin{lrbox}{\codeboxpim}
\scriptsize
\begin{lstlisting}[language=C,name=codeR,numbers=left]
// Executed on PIM-Enclave
char local_mem[BATCH_SIZE];
void *A_ptr;
get_params(A_ptr);

// Request data from bank memory 
dma_request(A_ptr, local_mem, 
    BATCH_SIZE, BANK_READ, DECRYPT);

// Perform computation on data
...

// Write the results into bank memory
dma_request(local_mem, A_ptr, 
    BATCH_SIZE, BANK_WRITE, ENCRYPT);
               
\end{lstlisting}
\end{lrbox}

\fbox{
\begin{minipage}{0.75\columnwidth}
    \usebox{\codeboxhost}
    \end{minipage}
}
\begin{minipage}{0.2\columnwidth}
\hfill
\vspace{1cm}
\end{minipage}%
\fbox{
\begin{minipage}{0.75\columnwidth}
    \usebox{\codeboxpim}
    \end{minipage}
}
\fbox{
\begin{minipage}{0.75\columnwidth}
    \usebox{\codeboxhost}
    \end{minipage}
}
\caption{An example program that offloads computation to \thename. The host processor enclave first performs attestation on PIM and loads the PIM local memory with a PIM kernel. By sending a command and parameters, the host invokes PIM to perform computation on the already inside the memory. }
\label{fig:flow}
\end{figure}

\begin{figure}[t]
\centering 
\includegraphics[ width=1\columnwidth]{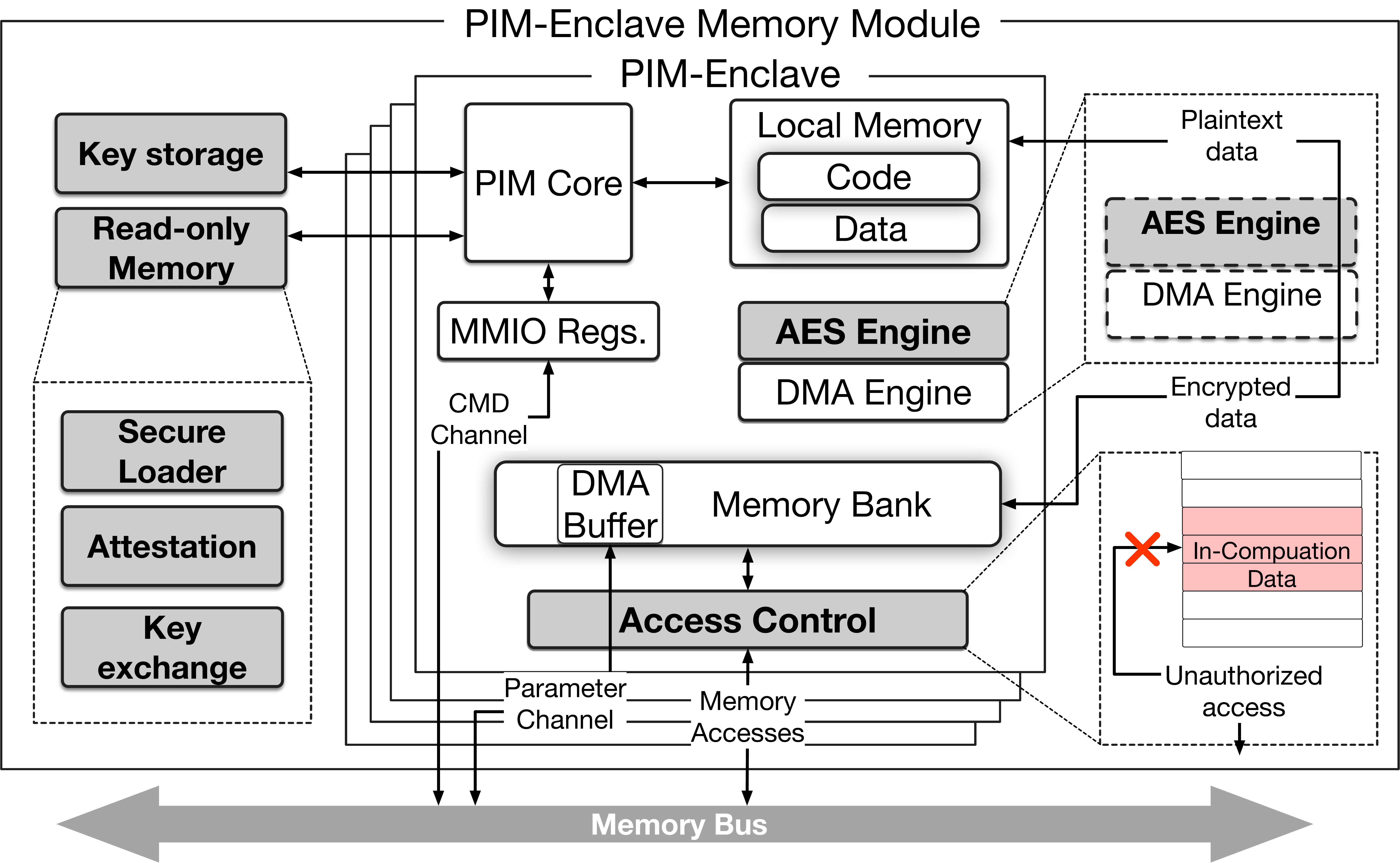}
\caption{An overview of \thename. Compared to the baseline, the architecture adds access control for memory banks and AES-capable DMA engine to ensure confidentiality and integrity of the computed data inside memory. The proposed modifications are colored in grey.}
\label{fig:components}
\end{figure}

\subsection{\thename Overview}
\label{subsec:overview}
We provide an overview of data computation acceleration using \thename at a high level before we proceed to explain the design in more detail. \autoref{fig:flow} illustrates a simplified example application for \thename.

\hdr{Offloading Model.} \thename constructs a confidential computing model in which the host-side enclave is assisted by the \thename for data-intensive computation, on top of the general PIM programming model. The host program acquires a handle (e.g., a file handle) and use it to communicate with the PIM core. Then, it performs attestation of \thename (line 2-3 in host-side code). The host program sends a PIM kernel, an executable compiled for PIM, to be loaded into PIM local memory by the PIM core. Then, a measurement of the PIM local memory is in order. Finally, the host sends a command instructing the PIM enclave to start execution.


\hdr{PIM Memory Model.} In \thename, a memory module is armed with compute-capable memory banks that can host secure enclaves. The memory bank serves as a conventional main memory for the host and storage for data during confidential computing. PIM kernels use PIM local memory to store intermediate results of computation. In our generalized PIM programming model, before the computation, the host load encrypted data to be computed upon into the memory bank~(line 8 in host-side code).
Then, during computation, the PIM core fetches a batch size of data from the memory bank to its local memory to perform computation via DMA (line 8 in PIM-side code). As we will explain further in \cref{subsec:memory}, the bank memory is \emph{locked} to thwart any access from the host and is only available to the PIM core (line 13 in host-side). 

\hdr{Zero-copy data passing. } Notably, there is no data copying between the host enclave and \thename during computation. After data is already located inside PIM memory banks, data passing between \thename and the host enclave in our programming model happens only through the passing of a pointer (line 16 in host-side code). The host passes a virtual-address pointer to the PIM core for it to translate to a bank address (e.g., through a fixed offset). Upon completion of the offloaded task, PIM simply notifies the host to retrieve the result inside memory or to request further computation on the data. The zero-copy passing of large data is a unique advantage of PIM-based accelerators in terms of performance. First, data travels to the cache through the memory bus only when the host access it. Second, the host can execute multiple iterations of the PIM kernel with different parameters or employ another PIM kernel to process data, all without accessing the data. We point out that such elimination of data movement during computation also eliminates the side-channel attack vectors. 

\hdr{Enabling Confidential computing with \thename .} Our \thename design retrofits the generalized baseline PIM design to support confidential computing. \autoref{fig:components} shows the overall design of \thename with the proposed modifications. Key storage containing the root endorsement key (EK) and ROM are added to service attestation and loading requests from the host enclave. As having ROM and key storage on each PIM core is expensive, we reuse them across different enclaves. We describe more about our support for remote attestation and secure channel establishment in~\cref{subsec:attestation}. \emph{AES engine} plays a crucial role in processing the encrypted in-memory data efficiently, as we will explain in \cref{subsec:aes-dma}. We also found that the ability to \emph{lock} access to a region of the memory bank while PIM core is processing data is imperative for confidential computing (\cref{subsec:memory}).

\subsection{Design Requirements for Confidential Computing Inside PIM}
\label{subsec:requirement}
We describe the requirements for supporting confidential computation in PIM by examining the attack vectors, propose modifications and also examine their performance impact on PIM.

\hdr{R1. Secure communication.} Establishing a secure channel between the processor enclave and a peripheral device is imperative in preventing eavesdroppers in the untrusted medium. The requirements for a secure peripheral device are well studied through many previous works~\cite{graviton,trustore,hix}; the device must actively participate in creating a secure channel. The specific hardware and software components necessary for achieving the goal are rather straightforward. They include secure key storage for the key that serves as a root-of-trust in authentication, capability to perform asymmetric key generation, and so forth. We adapted these well-defined requirements in \thename design.

\hdr{R2. Mitigation of side-channels for data in transit.} Memory bus transactions whose contents are encrypted can still leak information, since the adversary can monitor the bus packets to collect address access patterns. For processor enclaves, ORAM-based solutions~\cite{zerotrace,obfuscuro} can be considered if their high-performance overhead can be endured. The address side-channel on the memory bus is mitigated by design, since no data movement occurs on the memory bus \emph{during} computation. It should be noted that the existing PIM-based solution~\cite{invisimem,obfusmem} mitigate the address side-channel that occurs due to the ongoing computation inside the host enclave. As we will further explain in \cref{subsec:memory}, \thename brings computation inside memory and this requirement is satisfied by design.


\hdr{R3. Protection of data confidentiality and integrity.} Protecting data in memory during computation is a multifaceted issue. \thename proposes design changes that thwart unauthorized access and eliminate possible side-channels against multiple attack vectors.  Data can be stored in two places inside the memory module: the memory bank and the on-chip memory inside the PIM core package (PIM-local memory). The first attack vector is unauthorized access to the memory bank (\textbf{R3.A}). The adversary with a system privilege can access the physical address corresponding to the \thename memory bank currently in use. Encrypting the data with the shared key exchanged during initial attestation between the host enclave and PIM is sufficient for mitigating this particular attack¡. However, encryption alone does not mitigate side-channel, which leads us to the second attack vector: memory content change side-channel. Even if the memory content is encrypted, the attacker can monitor the memory content changes when the PIM outputs the computation result back to the memory bank in batch sizes (\textbf{R3.B}). Lastly, the memory module (e.g., the DRAM stick) can be subject to \emph{cold boot attacks}~\cite{coldboot-attack}, where the attacker dumps the contents of the memory (\textbf{R3.C)}.  We discuss protecting data confidentiality and integrity in \cref{subsec:memory}.

\hdr{R4. Performance overhead. } Besides the security requirements, we also found that working with encrypted data is too burdensome for low-power PIM cores. As we will present in \cref{subsect:aes-dma}, we evaluated the feasibility of a software-based AES on real PIM hardware and found that the throughput between PIM core and memory bank drops to less than 1\% when the PIM core performs DMA requests and AES cryptographic operations simultaneously. Hence, we concluded that hardware acceleration that does not consume PIM core cycles is essential for the \thename design.

\subsection{PIM Attestation and Secure Communication} 
In this section, we explain software and hardware components of \thename that help establish a secure communication channel between the host enclave and \thename (\textbf{R1} from \cref{subsec:requirement}).


\label{subsec:attestation}
\begin{table}[t]
    \footnotesize
    \centering
  	\renewcommand\theadfont{\bfseries}
    \begin{tabular}{ll}
        \toprule
        \thead{Command}&\thead{Description}\\
        \midrule
        \texttt{GET\_TOKEN}&Get attestation token\\ 
        \texttt{SET\_SESSION\_KEY}&Send encrypted session key to PIM\\ 
        \texttt{SET\_DATA\_KEY}&Send data decryption key to PIM\\ 
        \texttt{OFFLOAD\_KERNEL}&Ofload PIM kernel\\
        \texttt{EXECUTE}&Command PIM to execute the kernel\\ 
        \texttt{PROTECT}&Turn on the access control for a memory region\\
        \texttt{DESTROY}&Kill session and clear local memory\\
        \bottomrule
    \end{tabular}
    \caption{Commands supported by \thename}
    \label{tab:commands}
\end{table}

\hdr{Remote Attestation. }
The attestation scheme supported by \thename adapts the principles from the existing secure peripheral devices~\cite{graviton,invisimem,trustore,sdimm}. In particular, the device must support (1) secure key storage that contains the root endorsement key (EK), (2) self-attestation capability to establish the root of trust, and (3) the ability to generate asymmetric keys and perform a key exchange. The code that is used during attestation and initial loading of PIM kernels is stored in a ROM. The PIM cores share ROM and EK in key storage among multiple PIM cores in a memory module.  The attestation process for a processor enclave (i.e., SGX) and an external entity (e.g., a secure accelerator) is well-documented and provided with an ample amount of examples~\cite{sgx-ra,sgx-ra-tls,graviton,invisimem,trustore,sdimm}; therefore, we do not discuss it in detail in this work for brevity.  We implement the following commands for the attestation process:  \texttt{GET\_TOKEN} and \texttt{SET\_SESSION\_KEY} (\autoref{tab:commands}). Moreover, it is crucial to prevent a malicious enclave from leaking the EK, i.e., only allowing the attestation procedure to access the EK. As noted by \cite{minimalist}, the requirement can be easily enforced with an introduction of a hardware monitor on PIM core's program counter to the key storage that only allows access to the key when the PC is within the attestation code. 


\hdr{Securing communication between host and PIM. }
After attestation, the host enclave creates and shares a shared symmetric session key with PIM to establish an encrypted channel. The host enclave and PIM use the shared key to encrypt every traffic between them. The communication happens through two interfaces. The first interface is registers mapped into the host enclave memory space used to send commands for various operations~(the \emph{command channel}). The second interface is a shared DMA buffer for sending parameters for the commands~(the \emph{parameter channel}). The parameter channel employs a fix-sized buffer to prevent the side-channels incurred from packet varying sizes. The channel also employs the AES-enable DMA engine~(\cref{subsec:aes-dma}) for efficient parameters transfer and decryption between the DMA buffer and PIM local memory. The secure channel that uses AES-GCM also allows \thename to authenticate the sender of commands, preventing unauthorized entities from issuing commands to \thename. In addition to the secret session key, through the established secure channel, the host also shares a \emph{data encryption key} with \thename using the command \texttt{SET\_DATA\_KEY} to encrypt and decrypt data inside the memory bank. On receiving the exchanged keys, the PIM core programs them into the registers of the AES engine.  

\subsection{Protecting Data Confidentiality and Integrity}

\label{subsec:memory}

We explain the security measures for protecting data confidentiality and integrity on the memory bus and inside memory during computation. By doing so, we explain how our design satisfies the requirements \textbf{R2} and \textbf{R3\{A,B,C\}} described in \cref{subsec:requirement}). 

First and foremost, keeping data encrypted with the key shared by the host enclave and PIM protects the confidentiality and integrity of the data against unauthorized access. This prevents an entity that is not the host thread running inside the enclave from arbitrarily mapping and accessing the data. Encryption of data also prevents the plaintext from being visible in the memory bus to attackers that have placed physical probes on the bus. Having such relatively straightforward attacks put aside, now we discuss mitigation of side-channels during data transfer and computation.

\hdr{Side-channels in data-in-transit. } Computation inside \thename does not create any bus transactions during runtime. Therefore, confidential computing with PIM is inherently robust to the address side-channel on the memory bus (\textbf{R3.A}). Data movement only happens inside memory as the PIM core fetches a batch of data to be processed from the memory bank to its local memory. While an attack that probes the DIMM port exists~\cite{membuster}, probing the interconnects inside the memory module or on-chip local memory is not feasible under sensible circumstances, as the interconnection is often tightly integrated~(through stacking in 3D memories~\cite{invisimem} and per-bank integration in DIMM~\cite{upmem}). Also, we observe that the memory side-channels happen \emph{during} calculation due to a distinct pattern of behavior in the code. Therefore, placement of data in the memory bank \emph{before} computation and retrieving the data \emph{after} computation only leaves a sequential pattern with very little information to be leaked. Besides data,  commands to PIM and parameters (e.g., pointers to the data) may be passed as needed during computation. However, such transmissions are encrypted and fixed-size; hence, they do not leak any meaningful information to the adversary. On the other hand, we regard the side-channels that may arise due to computations on the host side. We assume that the general use case of \thename would be partial or full offloading of data-intensive and sensitive computation to the PIM side. We leave the mitigation of side-channels in the host-side computation to the programmer's discretion; the programmer may select sensitive parts of the program to run inside \thename or include mitigation such as ORAM for computations inside host enclaves.


\begin{figure}[t]
    \centering
    \includegraphics[width=\columnwidth]{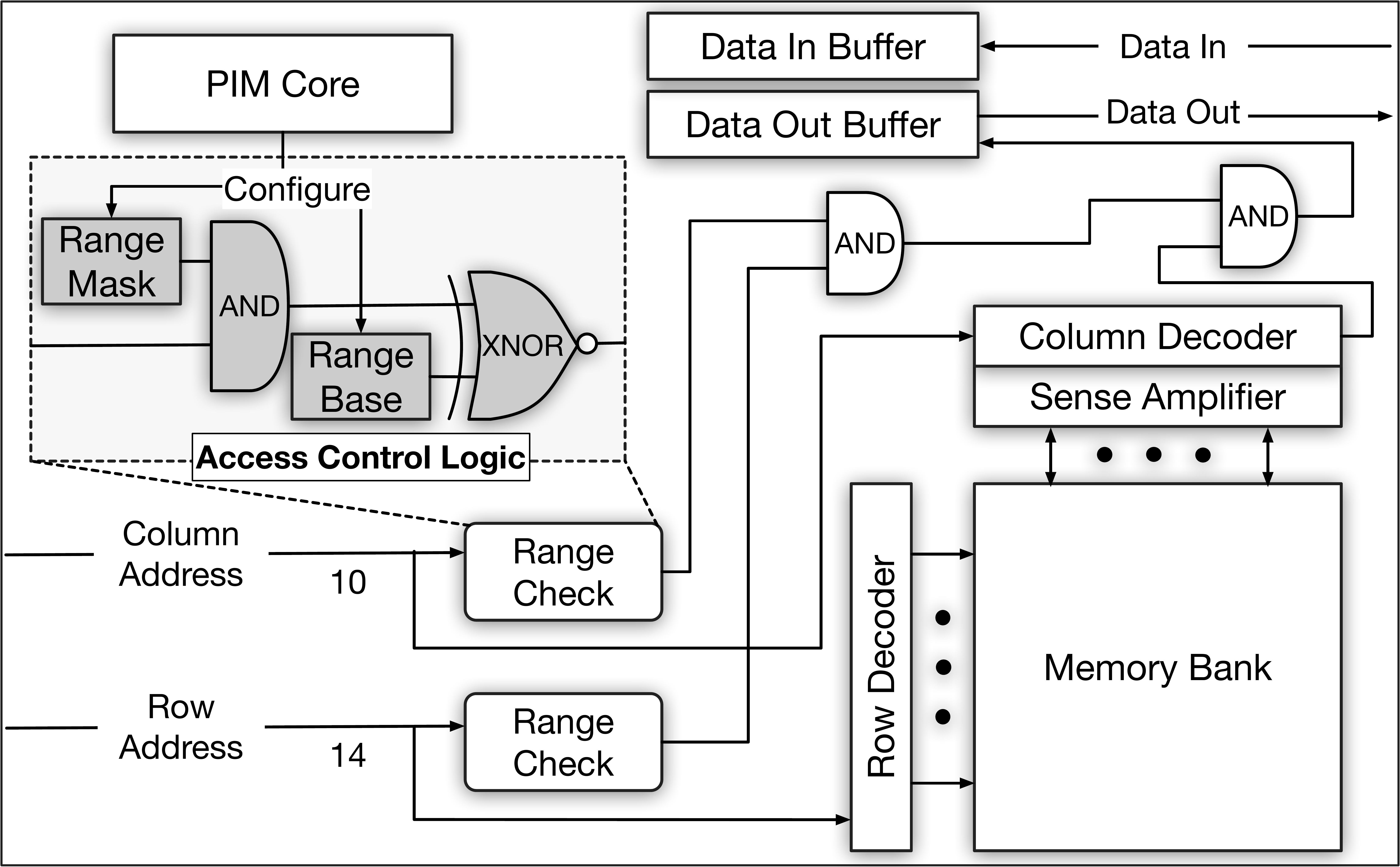}
    \caption{Access control logic in a DRAM bank}
    \label{fig:access-control-dram}
\end{figure}
\hdr{Protection of data during computation. } We concluded that an access control mechanism inside the memory module is essential to eliminate \emph{memory content change} side-channel for computation inside PIM. The main memory is shared by both the host enclave and \thename and can be used as a standard memory module (by mapping it into a virtual address space). Therefore, an untrusted entity in the host can observe the changes in the memory content even though it is encrypted. Due to the limited size of the PIM's local memory, the PIM kernel must work with the encrypted data in batches, which is not necessarily sequential. A batch of data will be fetched from the memory bank to be decrypted and processed. Then, the result will be encrypted and put back into the bank. Such operations may inadvertently create recognizable patterns in the eyes of the adversary. For this reason, we included an access control mechanism in \thename-enabled memory bank to thwart accesses from the host during computation (\textbf{R3.B}). 

\autoref{fig:access-control-dram} shows our proposed design  of the access control logic for DDR4 at the memory bank level. In the DRAM module, each incoming bus address is translated into a combination of \emph{bank group}, \emph{bank address}, \emph{row address} and \emph{column address}. Access control using the requested address can be enabled by introducing range checks for the column address and row address. In our design, the range-checking logic implements a pair of \emph{range mask} and \emph{range base} registers that are exposed to the PIM core for configuration. The incoming address is filtered  through the logic gates, and setting the range base and mask registers to \texttt{0x0} will disable the access control. Our design incorporates a similar circuit structure that already exists main processors for detect illegal memory accesses~\cite{sgx-explained}. 

We expect that the proposed access control logic can be incorporated into DDR4 DRAM modules, and 3D-stacked memory modules. We estimate that our range check logic, comprised of simple logic gates~(e.g., $10$ AND gates and $9$ XNOR gates to check 10-bit addresses), will introduce a negligible  latency. In a typical DDR4-3200 module, row activation takes approximately $15ns$ and column access, $24$ cycles.~\cite{micronddr4}. For 3D-stacked memory, the implementations often use packetized interfaces that replace the traditional low-level DDR commands. Hence, inserting the logic for access control is more straightforward. In fact, many works have taken advantage of the flexibility to customize packet handling logic~\cite{invisimem,hmc,obfusmem}. Therefore, we expect that our proposed logic change that implements access control can be easily incorporated into 3D-stacked PIM hardware.



\hdr{Consideration for cold boot attacks. } Our design choice was to keep the data residing in the memory bank encrypted, even with the new capability to drop access to the memory bank during computation. An adversary with physical access to the memory module may launch a cold boot attack to dump the memory contents in the memory bank. Therefore, to protect the contents of the memory bank, we need to employ cryptographic protection in parallel to maintain confidentiality even under such the worst case scenario (\textbf{R3.C}). This also implies that \textbf{R4} must be satisfied to achieve data protection. 




\subsection{Supporting Data-Intensive Computing with AES Engine}
\label{subsec:performance}
\label{subsec:aes-dma}

\begin{figure}[t]
    \centering
    \includegraphics[width=\columnwidth]{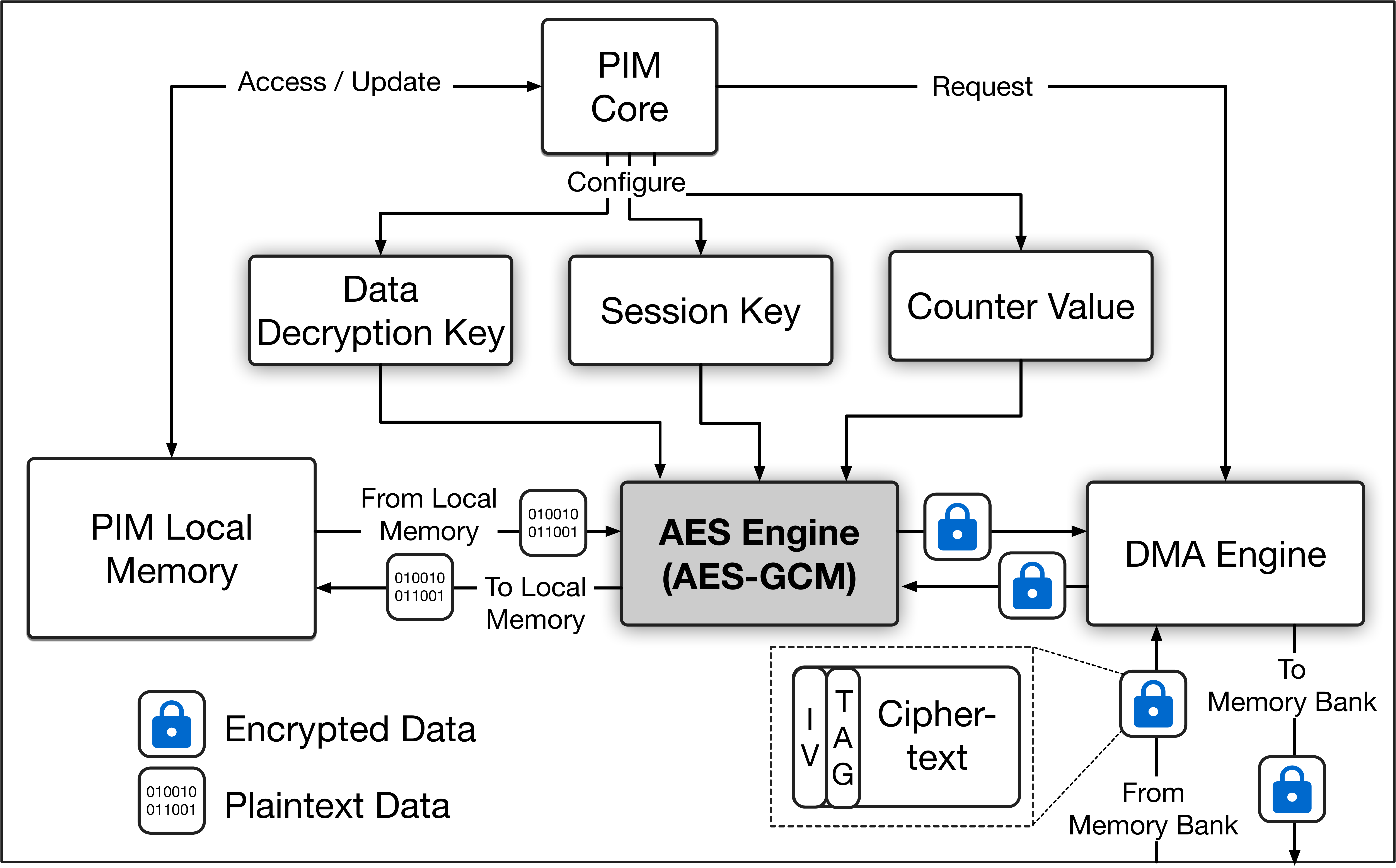}
    \caption{Structure of AES-capable DMA engine in \thename. Configurations for AES such as session key are exposed through registers to PIM core. DMA engine automatically performs encryption/decryption on DMA requests.}
    \label{fig:aes}
\end{figure}

\autoref{fig:aes} describes the AES-capable DMA engine and its operation. \thename proposes a hardware AES acceleration in the DMA engine of the PIM core to support computation on large and encrypted data. AES-enabled DMA would automatically encrypt or decrypt the data as the transfer occurs. In turn, the PIM core would configure the hardware with the shared key generated during secure channel establishment with the host enclave, and the data would arrive decrypted to the PIM local memory from the memory bank. As shown in our experiment with UPMEM in \cref{subsec:upmem-aes}, the currently available PIM hardware is not powerful enough to perform cryptographic operations and data computation simultaneously. We expect that the performance of the PIM core would not be drastically improved even in future iterations of PIM design and implementation such as HMC and HBM, as the power and area constraints will always be limiting factors in PIM design. An ISA extension for AES acceleration (e.g., x86 \texttt{AES-NI)} can improve the encrypted data transfer performance. However, we argue that it would consume a significant portion of the PIM core cycles that could have been used for data computation. Also, such a requirement creates an unnecessary constraint in PIM design. We propose the use of an AES acceleration capable DMA engine. Implementations of AES-enabled DMA engines already exists~\cite{nxp,advanced-dma-controller} and can be easily incorporated into existing hardware due to its predictable power and area requirements but can be an indispensable component in achieving confidential data-intensive computing.
For these reasons, we concluded that the inclusion of an AES engine is essential in our design to satisfy \textbf{R4}. We further explain the details of the component and our simulation methodology in \cref{subsec:impl-aes}.


\section{Implementation}
\label{sec:implementation}
We implemented a prototype of the \thename design using a cycle-accurate full system simulation~\cite{gem5}. As we explained in \cref{subsec:pim-baseline}, we generalized the PIM architecture and programming model. Regarding the low-level simulation configurations,  we referenced well-documented industry-grade products~\cite{upmem} and simulation configurations from other works on PIM ~\cite{smcsim,impica,obfusmem,ambit,tesseract}. Our simulation is well modularized and generalizable; the hardware components that we implement can be easily modified and provides well-defined APIs for writing host-side and PIM-side applications\footnote{We plan to make our simulation public after the publication of this work}. 




\subsection{AES-capable DMA Engine}
\label{subsec:impl-aes}

We implemented and incorporated the AES-GCM acceleration capability in the DMA engine  of the PIM core packages. Our strategy in incorporating a hardware design into the simulation was two-track: we first modified the DMA engines in the memory controller of the PIM core packages to include an AES-GCM acceleration hardware to test the correctness aspect. Then, we calculated the increased number of cycles consumed for each DMA transfer by inspecting the specifications of AES hardware specifications.

\vspace{5pt}
\hdr{Functional correctness.} We implemented AES functionality in the memory controller and its DMA engine of each PIM core. The implementation is based on open hardware IPs~\cite{aes-opencore,nxp} and uses functions from the \texttt{OpenSSL} library. We expose the configuration registers for various configuration parameters and command interface by mapping them into the PIM local memory. Through the command interface, the PIM core can specify whether the data should be encrypted or decrypted. Upon receiving a destination address from the PIM core, the memory controller initiate the transfer, then encrypts or decrypts the contents. The following is an example of AES-capable DMA usage in a PIM kernel:

\begin{scriptsize}
\begin{code}{C}
// Address inside memory bank
DMA_SRC_ADDR = bank_addr; 
// Address inside local memory
DMA_DST_ADDR = loc_mem_addr;
DMA_TRANSFER_SIZE = sz;
// Command the engine to decrypt the block
DMA_CMD = DMA_DECRYPT_TRANSFER;
poll(DMA_STATUS);
// Read the results
access(destination_addr);
\end{code}
\end{scriptsize}

Data is partitioned into encrypted blocks, each of which have an encryption IV and a tag appended to them for decryption. In particular, the size of each block should be

\vspace{2pt}
\begin{equation*}
\footnotesize
size_{block} = \frac{size_{available} - n_{split} * (size_{IV} + size_{tag})}{n_{split}},
\end{equation*}
\vspace{2pt}

where $n_{split}$ is the number of encrypted data blocks that fit into the local memory in one transfer. The developer can customize this value to optimize fetching and processing in \thename kernel. $size_{IV}$ and $size_{tag}$ are the sizes of IV and tag used in decryption and authentication, and $size_{available}$ is the maximum available size used to contain the data inside PIM local memory. Finally, the total number of encrypted blocks with respect to the data size is

\begin{equation*}
\footnotesize
n_{blocks} = \frac{size_{data}}{size_{block}}.
\end{equation*}
\vspace{1pt}

\hdr{Simulating latency. } The performance overhead of incorporating AES-GCM acceleration into DMA engines is well-studied and can be accurately estimated. We adapt the latency described in the specification of the IP~\cite{aes-opencore}. Also, previous works have estimated the latency of almost identical hardware~\cite{invisimem, obfusmem}. We estimated that the AES-GCM accelerator can operate at an $300Mhz$ clock rate and could encrypt 128-bit blocks each cycle. Based on the estimation, we added  add a delay of $1$ cycle to each 16 bytes data chunk processed by the DMA engine in our simulation.

\subsection{Memory Module Access Controller} 

We explain the modifications that enable temporary locking of memory banks in which confidential computation is in progress.  In our simulation, we made modifications to the \path{AbstractMemory} object in gem5, which is an abstraction of the memory controller. As \path{AbstractMemory} uses packets to interact with other components, its functionality is similar to how the logic layer of 3D-stacked memories handles the requests. In particular, the following logic is added to the \texttt{access()} function of \path{AbstractMemory} prevent read accesses into a protected region.

\begin{scriptsize}
\begin{code}{C}
void AbstractMemory::access(pkt) {
...
    if(pkt->isRead() {
        if (pkt->getAddr() >= protected_addr &&
            pkt->getAddr() <= protected_addr + 
            protected_size) 
        {
               set_empty(pkt); 
        }
        else serve_read(pkt); 
    }
...
}
\end{code}
\end{scriptsize}
We admit that it is rather challenging to simulate such a low-level mechanism in a simulator accurately, or even FPGA. We leave real hardware implementation of bank access control for DDR4 as future work.

\section{Evaluation}

\begin{table}[b]
    \footnotesize
	\centering
  	\renewcommand\theadfont{\bfseries}
	\begin{adjustbox}{width=1\columnwidth}
		\begin{tabular}{p{20mm}p{60mm}}
    		\toprule
			\multicolumn{2}{c}{\thead{Host Configuration}}                                   \\
			\midrule
			\thead{Processor} & 2 $\times$ in-order ARMv7 core (2 threads/core, clock rate 4.0GHz)\\
			\thead{L1 cache}  & Per-core, 32kB icache, 64kB dcache \\
			\midrule
			\thead{L2 cache}  & Shared across cores, 256kB\\  
			\hline
			\thead{Operating System} & \vspace{2pt} Linux kernel 3.16.0-rc6\\
			\bottomrule
		\end{tabular}
	\end{adjustbox}
	\vspace{1mm}
	\newline
	\begin{adjustbox}{width=1\columnwidth}
		\begin{tabular}{p{20mm}p{60mm}}
		    \toprule
			\multicolumn{2}{c}{\thead{PIM-enabled Memory Bank Configuration}}                                                               \\
		    \midrule	
			\thead{Processor}  & 1 $\times$ in-order ARMv7 core per bank (1 thread/core, clock rate $1GHz$)                             \\
			\midrule
			\thead{Local memory} & $4MB$ per core, access latency $0.01ns$\\
			\midrule
			\thead{Memory Bank} & 64MB/bank, Memory type: \texttt{HMCVault} in \texttt{gem5}, 8 $\times$ \thename-enabled bank \\
			\bottomrule
		\end{tabular}
	\end{adjustbox}
	\vspace{1mm}
	\newline \begin{adjustbox}{width=1\columnwidth}
		\begin{tabular}{p{20mm}p{60mm}}
			\toprule
			\multicolumn{2}{c}{\bfseries DRAM Configuration}                        \\
            \midrule
			\bfseries DRAM parameters & Row buffer size: 256B, burst size: 32B
			tRP: 13.75ns, tRCD: 13.75ns, tCL: 13.75ns, tBURST: 3.2ns                    \\
        	\bottomrule
		\end{tabular}
	\end{adjustbox}
	\caption{Configuration of the simulated system.}
	\label{table:config}
\end{table}

\begin{figure*}[!ht]
  \centering
  \begin{subfigure}{\columnwidth}
  \includegraphics[width=\linewidth]{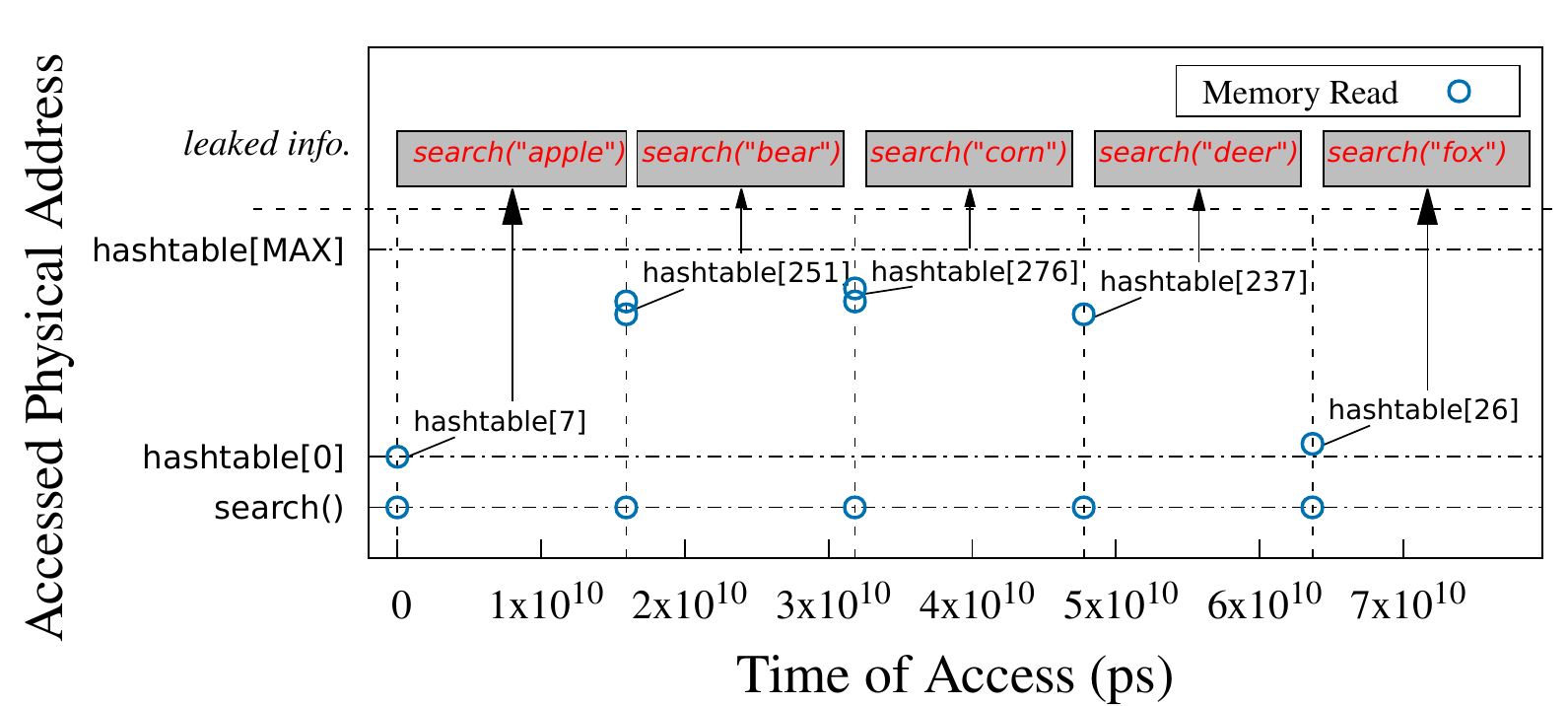}
  \caption{\hj{Memory access pattern observed on the memory bus during host processor enclave dictionary lookup.}}
  \label{fig:host-pattern}
  \end{subfigure}
\begin{subfigure}{\columnwidth}
  \centering
  \includegraphics[width=\linewidth]{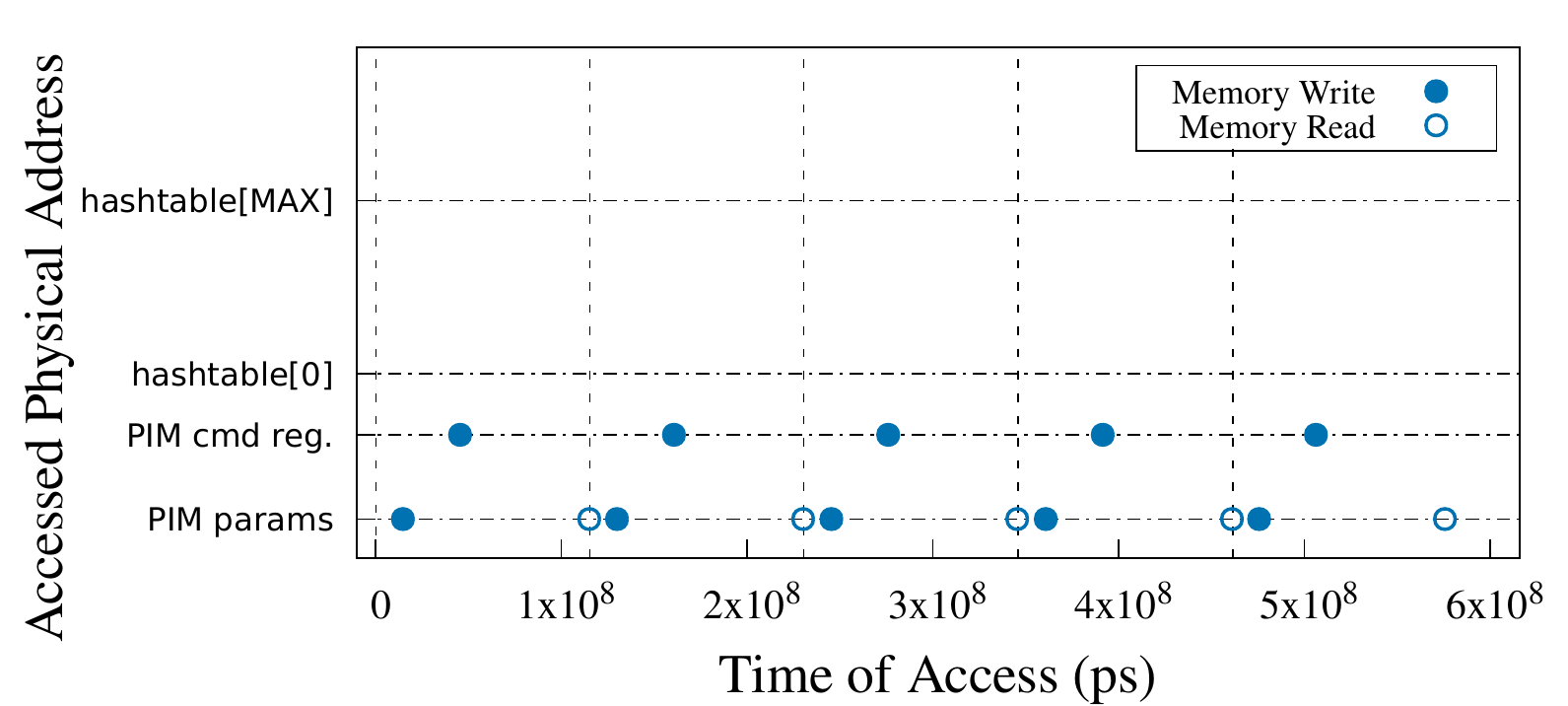}
  \caption{Memory access pattern observed on the memory bus during PIM-assisted dictionary lookup.}
  \label{fig:pim-pattern}
  \end{subfigure}
  \caption{\hj{Memory access pattern exposed on the bus of CPU-based hash table access compared to PIM-based hash table access. The vertical lines mark the timestamps in which the access function are called.}}
  \label{fig:access-pattern}
\end{figure*}

We evaluated \thename through a cycle-accurate simulation. The simulation configuration we used for simulating our system can be found in \autoref{table:config}. Note that we assume the presence of processor enclaves in the host, as our simulator does not support simulation of enclaves such as SGX~\cite{sgx} or TrustZone~\cite{arm-tz}. That is, we assume that \thename cooperates with the host enclaves to achieve confidential computation. For this reason, we did not consider the overhead from the use of host processor enclaves, although the exclusion of the overhead works against our favor. We configured the PIM core to have one core that operates at the clock rate of $1GHz$ to reflect the power and area constraint in the DRAM or 3D-stacked memory packages. The timing parameters used for the simulation were determined through the existing works that conducted simulations with similar hardware~\cite{smcsim,concurrent-data,tesseract}. There are a total of 8 \thename-enabled memory banks inside the memory module that can run in parallel.

We first illustrate the elimination of bus traffic side-channel in \cref{subsect:memory-access-pattern}. We provide an analysis of software encryption overhead on UPMEM, a real PIM hardware in \cref{subsec:upmem-aes} to motivate the necessity of our AES-capable DMA engine design. The performance of encrypted data movement inside \thename using the AES-capable DMA engine is shown in \cref{subsect:aes-dma}. Lastly, we evaluate the overall computation performance of \thename-assisted k-mean clustering application in \cref{subsec:app-perf}.

\subsection{Memory Access Pattern Analysis}
\label{subsect:memory-access-pattern}

\autoref{fig:access-pattern} illustrates the memory access pattern observed on the bus -- to the attacker who is snooping on the bus traffic -- in host processor only execution and \thename-assisted execution. We shall The experiment imitates the off-chip bus snooping attack illustrated in Membuster~\cite{membuster}. The attack demonstrates that bus snooping can undermine the confidentiality of processor enclaves through memory access pattern analysis on the bus traffic. 
Our test program is a minimal hash table lookup that mimics the attack examples on the \texttt{hunspell} dictionary program shown in Membuster. 

Each entry in the hash table stores a hash key and an integer value in the place of the word's definition for simplicity. The \texttt{search()} function  takes a string as an argument. The argument is hashed using the \emph{Murmur3} hash algorithm, then used as an index to the hash table. The exact operation of \texttt{search()} is as the following:

\begin{scriptsize}
\begin{code}{C}
struct data_item *search(const char* key) {
    int index = murmur3_hash(key) % TABLE_SIZE;
    while (hashtable[index].value != -1) {
        if(key_match(hashtable[index], key)){
          return &hash_array[index];
        }
        ++index;
        index %= TABLE_SIZE;
    }
    /* Key not found */
    return NULL;
}
\end{code}
\end{scriptsize}


The original attack from~\cite{membuster} requires that the host enclave caches are constantly flushed to yield reliable and distinct access patterns on the bus. We adapted the same premise for our experiment. However, since the gem5 x86 CPU model does not correctly simulate cache flush instructions (e.g., \texttt{clflush}), we implemented a function that forces cache flush by repetitive memory accesses and used it after each hash table access to preventing caching. 

In the host-only setup, which represents a case where the host processor enclave is used to protect the hash table, we place the hash table inside a single memory page so that it is contiguous in physical memory. Then, we obtain the virtual to physical mapping of the program from \texttt{/proc/\{process pid\}/pagemap}. 

In the PIM-based setup, \hj{For the PIM-assisted version of the program, the hash table is stored in the PIM memory bank, and a PIM kernel that performs hash table lookups is loaded into the PIM processor local memory.}  We leverage our simulation framework to place a memory tracing unit between the L2 caches and DRAM (i.e., the memory bus ) to collect the accessed \hj{physical addresses}.

\autoref{fig:host-pattern} shows the memory access pattern recorded during the execution of the host-only version of the test program, and \autoref{fig:pim-pattern} shows that of the \thename-assisted version. The memory writes are denoted by $\bullet$ and reads by $\circ$. 

 In the host-only version, the executed function and the words looked up in the hash tables are observable to the adversary. With an assumption that the adversary has the pre-obtained knowledge on the addresses of the functions and data of the program, the adversary learns that \texttt{search()} has been executed and in-memory dictionary locations, such as \texttt{hashtable[7]}, \texttt{hashtable[237]}  are accessed. As a result, the adversary can eventually infer the queried words by looking at the bus traffic. 

On the other hand, the PIM-assisted version shows only the communication with PIM. The PIM command channel and parameter channel are fixed single-channel memory-mapped I/O channels. The host program first writes the parameter (e.g., \texttt{"appple"} to the parameter channel, then writes the command number that corresponds to the PIM function \texttt{search()} in the PIM kernel to the command channel to invoke the offloaded task. These patterns are identical to all PIM kernel invocation, regardless of the invoked function and parameters. Also, data access (e.g., the hash table in this example) is contained inside the DRAM package. This example clearly shows the security advantages of \thename against side-channel attacks that target the bus traffic. We further discuss the security of \thename in a more comprehensive manner in \cref{sec:discussion}.

\begin{figure}[hb]
    \centering
    \includegraphics[width=\columnwidth]{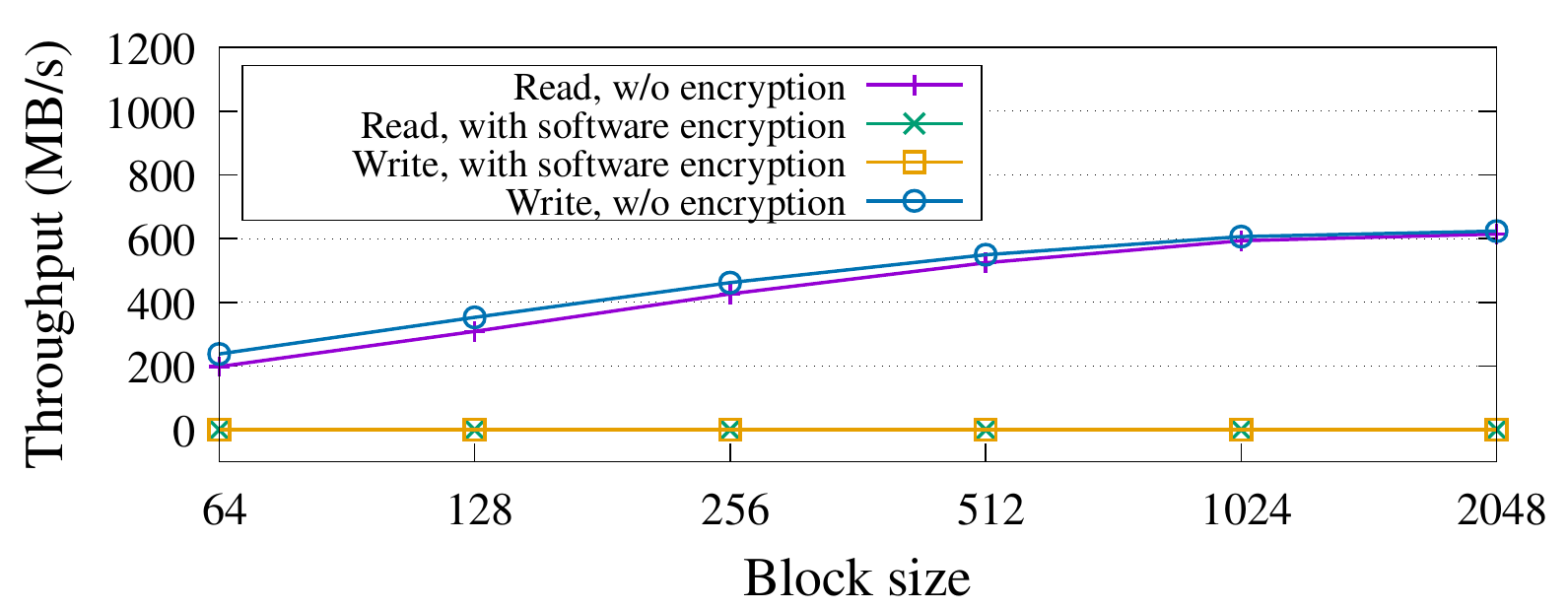}
    \caption{The memory throughput with and without software AES encryption on UPMEM PIM hardware.}
    \label{fig:upmem-aes}
\end{figure}

\begin{figure*}[ht]
  \begin{subfigure}{.25\textwidth}
  \centering
  \includegraphics[width=\linewidth]{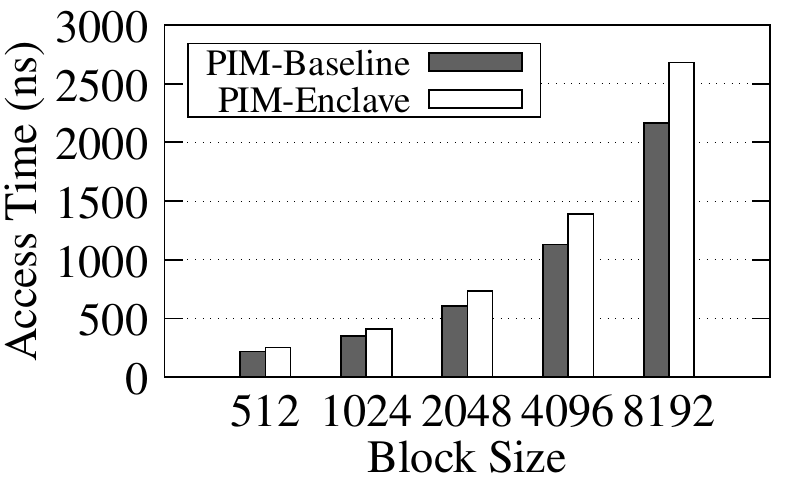}
  \label{fig:sfig1}
   \vspace{-1em}  
  \caption{Sequential read}
\end{subfigure}%
  \begin{subfigure}{.25\textwidth}
  \centering
  \includegraphics[width=\linewidth]{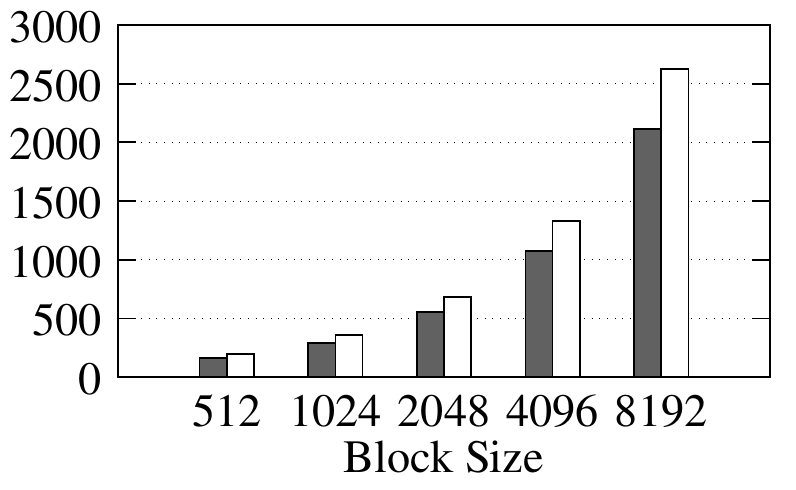}
  \label{fig:sfig1}
   \vspace{-1em}
  \caption{Sequential write}
\end{subfigure}%
  \begin{subfigure}{.25\textwidth}
  \centering
  \includegraphics[width=\linewidth]{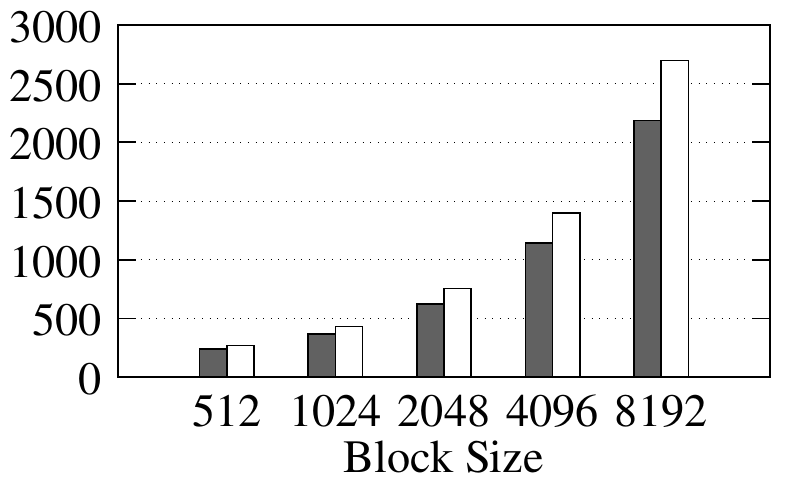}
  \label{fig:sfig1}
  \vspace{-1em}
  \caption{Random read}
\end{subfigure}%
  \begin{subfigure}{.25\textwidth}
  \centering
  \includegraphics[width=\linewidth]{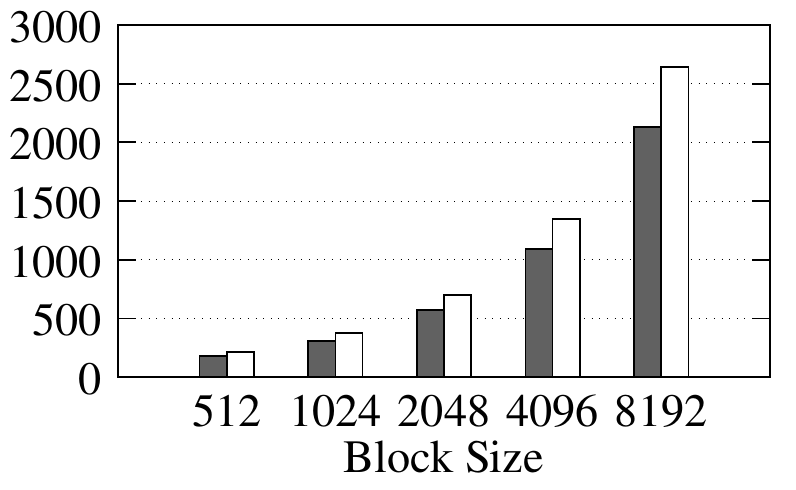}
  \label{fig:sfig1}
   \vspace{-1em}    
  \caption{Random write}
\end{subfigure}%
\caption{Access latency of AES-encrypted DMA from bank memory to PIM local memory of \thename vs. PIM-Baseline without encryption}
\vspace{1em}
\label{fig:dma-latency}
\end{figure*}



\subsection{Software AES on Real PIM Hardware}
\label{subsec:upmem-aes}

As shown in \autoref{fig:upmem-aes}, we evaluated the performance of working with encrypted data using software-based AES using \emph{mbedtls} on UPMEM's PIM hardware~\cite{upmem}. We measured the performance of encrypted DMA transfers both ways; the PIM core encrypted the data before transferring it from its local memory to the memory bank and decrypt the data as the data comes in the reverse direction. We measured the latency of each transfer from the initiation of encryption/decryption to the completion of the DMA transfer. The software AES incurs an immense performance overhead as shown in our experiment. The throughput plummeted from 600 MB/s to 0.5 MB/s, around 1200$\times$ lower throughput for both read and write. The results convinced us that hardware acceleration for symmetric key cryptography must be included for confidential computing in PIM design. Unlike convention accelerators (e.g., GPU), PIM designs have more constraints in terms of power and area constraints. We expect that even future generations of PIM hardware such as HBM~\cite{amd-hbm} will have limitations as to increasing the performance of the PIM core, and a hardware-based acceleration for cryptography will be an essential part of the design.

\subsection{Encrypted Data Movement Performance}
\label{subsect:aes-dma}

We evaluated the performance of encrypted data movement using \thename simulation to measure its performance. We measured the access times and throughput of AES-GCM-enabled DMA transfers in \thename against the baseline PIM simuation model, which we shall call \emph{PIM-Baseline}, that does not support any cryptographic protection for DMA transfers. Additionally, we measured the performance of encrypted data transfer using software-based AES on UPMEM's real PIM hardware, and the result from the experiment motivated us to include the AES acceleration in our design.

\hdr{\thename access time. } For both PIM-baseline and \thename, we measured the access time 1. in both directions (local-to-bank and bank-to-local), 2. sequential and random accesses and 3. for varying block sizes. \autoref{fig:dma-latency} illustrates the average access time measured for $1000$ DMA accesses in each configuration. Note that the access time for random and sequential access is roughly similar because there is no cache on PIM that incur variations in access time~(explained in our baseline model~\cref{subsec:pim-baseline}). Our simulation results report an average of $22.35\%$ of increase in access time due to the AES-capable DMA engine. This is in line with the previous works that simulated the performance overhead of hardware AES support such as~\cite{obfusmem,invisimem}.

\hdr{\thename throughput. } We further investigated the throughput for data transfer between the memory bank and local memory. The DMA maximum throughput of moving data with the AES-capable DMA engine is compared against that of without encryption in PIM-Baseline. Overall, the AES-enabled DMA engine produces an average throughput of 2.9 GB/s, compare to 3.53 GB/s of unencrypted DMA accesses.


\begin{algorithm}[b] \scriptsize
    \begin{algorithmic}[1]
    \Procedure{update\_clusters}{$I$,$O$,$M$,$delta$}
    \For{Object $obj$ in $O$}
    \State $cluster \gets findNearestCluster(obj,I)$
    \If{ $M[obj]$ is not $cluster$  }
    \State $M[obj] \gets cluster$ 
    \State Increment $delta$
    \EndIf
    \EndFor
    \EndProcedure
    \Procedure{compute\_kmean\_clusters}{}
    \State $I \gets PARAMS.CLUSTERS$
    \State $P_{Objects} \gets PARAMS.OBJECTS$
    \State $P_{Membership} \gets PARAMS.MEMBERSHIP$
    \State $delta \gets 0$
    \Repeat
    \State Get objects $O$ from $P_{Objects}$
    \State $O \gets DMA\_READ(P_{Objects})$ 
    \State $M \gets DMA\_READ(P_{Membership})$ 
    \State $UPDATE\_CLUSTERS(I, O, M, delta)$
    \State $DMA\_WRITE(M, P_{Membership})$ 
    \State Increment $P_{Objects}$,$P_{Membership}$ 
    \Until{All objects have are processed}
    \State Send size of clusters, $delta$, $I$ to host
    \EndProcedure
    \end{algorithmic}
    \caption{$k$-mean clusters update on \thename}
    \label{alg:kmean}
 \end{algorithm}

\subsection{Application Performance Analysis}
\label{subsec:app-perf}
We developed a proof-of-concept application to evaluate the performance in data-intensive applications and the feasibility of the programming model. We implemented a $k$-mean clustering application into a PIM kernel that runs inside \thename. $k$-mean clustering is a standard algorithm in data analytics that aggregates $N$ data points into $k$ clusters, where the data point belongs to the cluster with the nearest mean. The algorithm starts by selecting a group of $k$ coordinates as the \emph{centroids} of the initial clusters, then iteratively updates the centroids of clusters in two steps. First, it \emph{assign} every data point to their nearest cluster, i.e., the cluster that has the smallest \emph{Euclidean distance} to the data point. Then, it \emph{update} the centroids by finding the mean of data points that belong to a cluster.

\hdr{Secure $k$-mean clustering algorithm. }  In our $k$-mean implementation, we offload the core algorithm that incurs frequent memory access to \thename. The host distributes the encrypted data into multiple \thename memory banks, sends commands and parameters to \thename PIMs to initiate computation, then collects and aggregates the results into new \emph{global centroids}.  Upon receiving the command, the $k$-mean PIM kernel is executed in response. The kernel finds \emph{pointer} to the data to be computed in the DMA buffer along with other parameters. Algorithm \autoref{alg:kmean} shows the algorithm for updating centroid inside PIM.

We preprocessed the dataset used in $k$-mean by partitioning and encrypting the data coordinates into blocks of $8KBs$. We then prepend the block with the encryption IV and tag to be used for decryption. When applied to the chosen the dataset (the \emph{letter} dataset from \cite{uci-repo}), the preprocessing step results in blocks of encrypted data with 127 objects and 16 data points for each object. For each encrypted data block, we also use a structure to store the \emph{membership data} (e.g., which cluster the objects belong to) of objects in the block and offload them along with the objects into the memory bank.
The kernel used for secure $k$-mean performs $k$-mean update on encrypted data inside of memory bank in a batch of $127$ objects at a time. Using the AES-enabled DMA engine, it fetches the objects and their membership data into local memory. It performs the update procedure on the batch (\texttt{UPDATE\_CLUSTER} in \autoref{alg:kmean}) and keeps the intermediate results in the local memory. Finally, the kernel encrypts and updates the new membership data into memory before moving on to the next batch.

\hdr{Security Implications.}
In our implementation of $k$-mean, no data movement occurs between the host enclave and PIM between each execution of the kernel. Instead, only the parameters for the kernel, which includes the pointer of data, the size, the initial centroids, are sent from the host enclave to PIM. After a PIM kernel finishes its execution, only computation results are sent back to the host enclave. The elimination of data movement has two significant implications. First, efficiency was achieved in terms of the total execution time of the particular task and memory bus bandwidth usage.  Second, the computation did not leave any pattern on the memory bus. This is only possible on PIM-based computation since CPUs or any accelerators on the peripheral bus inevitably access memory with a pattern unless compiled with ORAM primitives. 

Also, our host enclave programs send the \texttt{PROTECT} command to lock the memory bank to which the computation will be offloaded. This means that all access to the memory bank will be simply dropped until the completion of the offloaded PIM kernel that will unlock it just before its termination. Hence, unauthorized entities cannot observe the memory content changes while the computation is in progress.

\begin{figure}[!t]
  \setlength{\belowcaptionskip}{-10pt}
  \centering
  \includegraphics[width=0.95\linewidth]{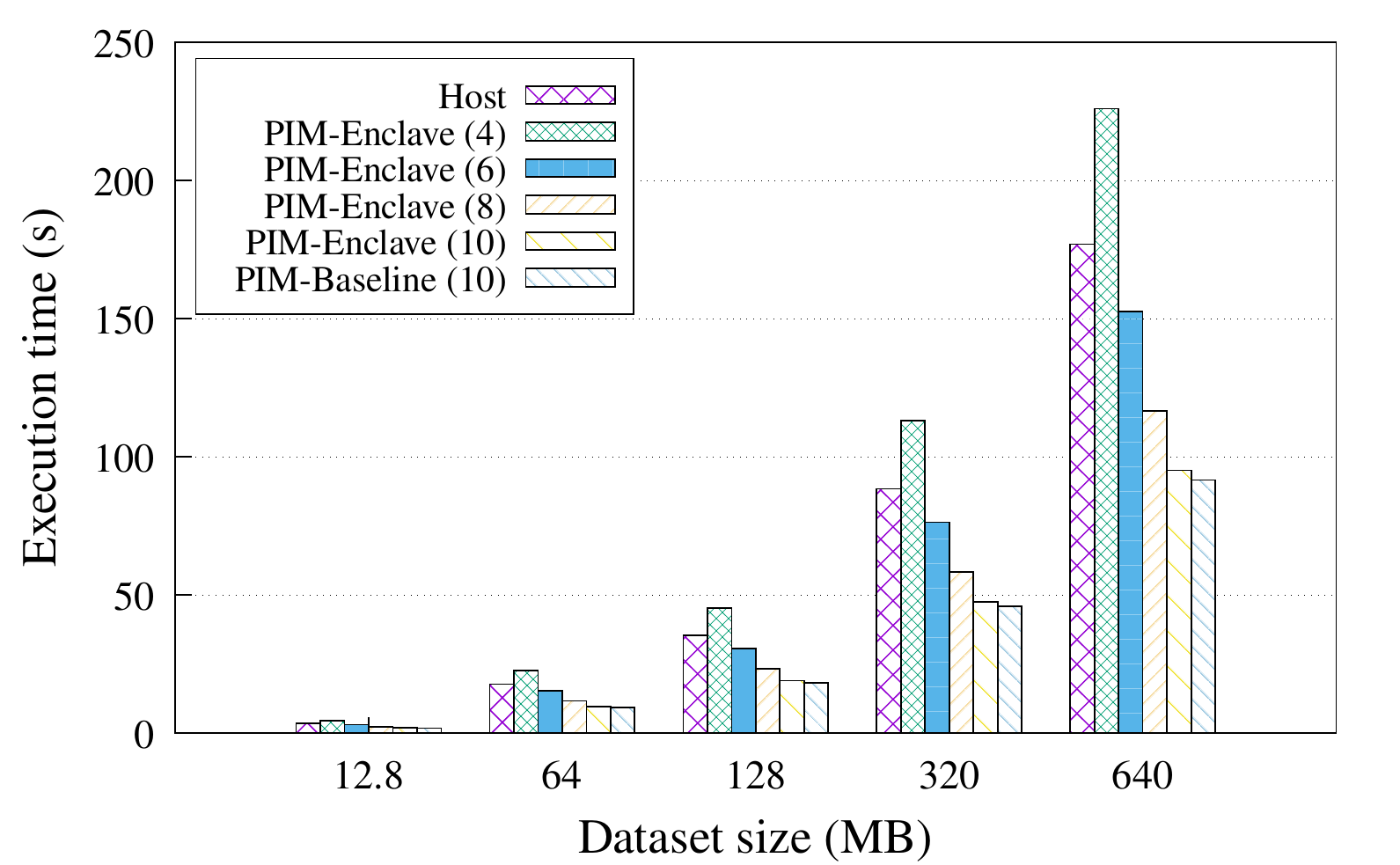}
  \caption{Performance of $k$-mean on PIM-Enclave against the host and baseline PIM (without data encryption). Data is projected up to $640MB$.}
  \label{fig:kmean-perf-versus-baseline}
\end{figure}


\hdr{Execution time of secure $k$-mean. }
\autoref{fig:kmean-perf-versus-baseline} demonstrates the performance of our implementation of $k$-mean on \thename in comparison to executing the same algorithm on the host and the baseline PIM. Note that in the host and baseline PIM case, data is not encrypted for a fair comparison. The reason is that gem5 does not support AES acceleration, and software encryption incurs significant overhead. For the baseline PIM, we only show the execution time when using 10 PIM devices. To shorten the simulation time,  we only evaluate the data set up to $1.28MB$, which has corresponds to $20,000$ data points and project the execution time of data based on the previous results. This is reasonable because, through our experiment, we found that the execution time of the algorithm scales linearly with the more data it process. The algorithm is repeated for $20$ rounds. The simulation results demonstrate that using \thename as an accelerator for secure data-intensive computation is efficient. It achieves a low overhead compared to the baseline model without encryption. When only $4$ \thename PIM devices are deployed, the execution time of $k$-mean is below that of the host on the same amount of data, with $21.7\%$ performance degradation. The reason is that there is a bottleneck in the communication between the host and PIM. However, \thename out-performed the execution time of the host with more than six \thename-enabled PIMs. Overall, the execution time of the secure $k$-mean version operated on encrypted data incurs only $3.7\%$ overhead compared to the insecure version of the algorithm that operates on unencrypted data. 

\section{Discussion}
\label{sect:sec-anal}
\label{sec:discussion}


\begin{table}[b]
    \footnotesize
    \centering
  	\renewcommand\theadfont{\bfseries}
    \begin{tabular}{p{3.0cm}p{5.0cm}}
        \toprule
        \thead{Attack Vectors} & \thead{Mitigation}\\
        \midrule
        Bus Traffic Side-channel & 
         No data movement  \newline
         Programming model \newline
        Secure communication channel~(\cref{subsec:attestation}) 
        \\ 
        \midrule
          Cold boot attacks & Maintaining bank memory encrypted during computation  ~(\cref{subsec:memory})  \\ 
            \midrule
        Unauthorized encrypted bank memory content accesses & Memory access control mechanism~(\cref{subsec:memory})\\

        \bottomrule
    \end{tabular}
    \caption{Attack Vectors on \thename's confidential computation and mitigations.}
    \label{tab:sec-analysis}
\end{table}

\subsection{Security analysis}

\autoref{tab:sec-analysis} summarizes the possible attack vectors and our proposed mitigation. As we demonstrated in our evaluation (\cref{subsect:memory-access-pattern}), the bus side-channel is eliminated by design in the \thename model. We incorporated an AES-capable DMA such that the bank memory content can be maintained in encrypted form without high performance overhead, and by doing so we prevent cold boot attacks. Our access control logic that filters incoming addresses to memory during PIM computation to prevent the unauthorized system software (e.g., kernel privilege) from monitoring the encrypted memory content changes. 

\thename provides a confidential computation offloading model that the host enclave can leverage to isolate and accelerate specific workloads securely. Nevertheless, the host program must be written with caution to achieve complete confidentiality. That is, the side-channels in host-side computation persists. Therefore, the developer can choose to offload all sensitive workloads to \thename or use host enclave side-channel mitigation (e.g., enclave ORAM~\cite{zerotrace}) when host-side computation must be mixed.



\subsection{Potential Side-channels Inside Memory Module}
There is currently no means of communication or shared resource among the memory banks inside \thename-enabled memory module; a \thename core can only access its local memory and its bank memory (i.e., it cannot access other bank memory). Also, at most, one host enclave can establish a secure channel and use a \thename bank. Such strict separation prevents security risks from malicious co-tenants who also have access to one or more \thename banks. There are no hardware connections that can be exploited for creating side channels. Being the first to explore confidential computing inside PIMs, our design focuses on maintaining the confidentiality of the computation. However, we expect that introduction of secure communication between PIM banks can enable more flexible forms of computation inside PIM. However, we leave such computations model as future work.

\section{Related Work}

\subsection{Processing-In-Memory}
Processing-in-memory (PIM) is the computing paradigm that integrates lightweight computing logic into memory, thus alleviating some computation from the central processor. PIM architecture has not been generalized, and the industry and academia propose different architectures. Most architectural researches on PIM focus on integrating the acceleration logic into the logic layer of 3D-stacked memories such as HMC or HBM~\cite{top-pim,tesseract,lazypim,pei,smcsim,graphpim,samsung-pim}. Most PIM research PIM on traditional DIMM focuses on placing fixed-function logic inside the DRAM circuit to perform some operations within memory~\cite{sdimm,ambit,newton,gradpim}. Recently, UPMEM~\cite{upmem} announced their first commercially available PIM architecture that embedded processing cores into the memory banks of DDR-based memory. 

\hdr{PIM-Accelerated Applications. } 
Due to its nature of being near memory, applications offloaded to PIM gain a high memory bandwidth as they do not have to move data across the slow memory bus. Moreover, in 3D-stacked memories, TSV connection between the layer naturally provides more internal bandwidth~\cite{primer-on-pim}. This makes PIM-based applications exceed memory-bounded workloads and workloads with erratic memory access patterns. Graph processing workloads~\cite{impica,graphpim,tesseract,pointer-chasing} and machine learning workloads~\cite{newton,prime,natsa,googleworkloads,massively-parallel-skyline,neural-network-training} are two of the most common use cases of PIM. Computational genomic is another application that requires the high internal bandwidth that is explored in several related works~\cite{upmem-dna-mapping,blast-upmem,medal}.

\hdr{PIM for security. } InvisiMem \cite{invisimem}, and ObfusMem \cite{obfusmem} are one of the first to harness processing-in-memory for security by instrumenting the logic layer of a smart memory module with cryptographic primitives to achieve security guarantees similar to Oblivious RAM. In particular, they enhance the host-side memory controller to authenticate with the smart memory and encrypt all packets sent to memory. Other works demonstrate that PIM is beneficial for accelerating security operations \cite{sdimm, near-data-security, hega}. \cite{near-data-security} suggests delegating Bonsai Merkle Tree memory authentication operations to memory while Secure DIMM (SDIMM) \cite{sdimm} offloads that a part of the oblivious RAM (ORAM) access routine to processing-in-memory. In \cite{hega}, the authors propose HEGA, a PIM architecture to accelerate homomorphic encryption search. All of those researches demonstrate that PIM can greatly reduce the memory bandwidth of memory-intensive security applications. 

Unlike \cite{invisimem,obfusmem,sdimm}  \thename does not require host architecture modifications. In particular, \cite{invisimem, obfusmem} modify the host memory controller to perform authenticated encryption with the smart memory module. \cite{sdimm} extends on Freecursive ORAM \cite{freecursive-oram}, a secure processor architecture that supports ORAM primitives. In contrast, our work can integrate with existing processors.
Our work further advances the aforementioned approaches by presenting a design for a general-purpose PIM enclave, which makes PIM available for a wide range of security applications.

\subsection{Confidential Computing In Cloud}
Supporting confidential computing in untrusted cloud a common goal shared by the industry and academia, as evidenced by confidential computing services being offered by cloud service providers \cite{azure,google-confidential-computing,nitro}. A number of works have proposed various SGX-based confidential computation schemes. The works explored the design space in using Intel SGX for secure data computation in the cloud~\cite{securekeeper,vc3,enclavedb,occulemency,opaque,oblivious-ml}. VC3~\cite{vc3} proposes an architecture for performing distributed MapReduce in untrusted cloud in which sensitive code and data are protected. EnclaveDB~\cite{enclavedb} protects the database and the querying process using SGX. Occlumency~\cite{occulemency} proposes an SGX-based deep learning inference that ensures confidentiality and integrity of user data.

The aforementioned works that employ SGX for secure data computation had to be designed considering the limited memory capacity of Intel SGX. SGX supports up to 128MB of \emph{Enclave Page Cache (EPC)} that is used for storing protected code and data. \cite{enclavedb} had to evaluate their performance with large databases with simulation due to the memory limit. \cite{occulemency} devised a batching scheme to perform deep learning inference with limited memory. Our work explores a design space with memory equipped with PIM participates in confidential computing with processor enclaves. Our model, \thename, assists the processor enclave in storing large data securely in memory and also allows offloading of data-intensive computations into memory.

\subsection{Side-Channel Attacks and Defenses in Processor Enclaves} 
\hdr{In-system side-channel attacks.} 
A number of previous works have shown that the confidentiality guarantees of processor enclaves (i.e., Intel SGX) can be undermined through various side-channel attacks. 
Many previous works have shown many cache side-channel attacks also apply to SGX\cite{leaky-cauldron,sgx-cache-attack,sgx-grand-exposure}. Controlled side-channels can arise due to the dependency of SGX on the untrusted kernel. \cite{sgx-page-table} exploits page faults on enclave pages to profile the memory access pattern of enclave programs. \cite{sgx-step} has shown the SGX enclave execution can be single-stepped through manipulation of APIC timer interrupts. 

\hdr{Bus side-channel attacks.} 
Besides the in-system side-channels, code and data in transit to memory can reveal enclave-protected program behavior. SGX only stores encrypted data in memory, however, a powerful adversary who can probe the memory bus transactions can collect the accessed addresses to reveal access patterns~\cite{membuster}. 

\hdr{Side-channel Mitigations} Most side-channel attacks aims to leak the memory access pattern of programs to leak sensitive information~\cite{leaky-cauldron,sgx-cache-attack,sgx-grand-exposure,sgx-page-table}. Transforming code and algorithm into a data-oblivious form is a mitigation method explored in the literature~\cite{raccoon,oblivious-ml}. Oblivious RAM (ORAM)~\cite{ring-oram,freecursive-oram,computedram} is accepted as a general solution for the access pattern side-channel and has been adapted for SGX~\cite{zerotrace,obfuscuro}. However, the performance overhead of ORAM-based approaches may deter their adaption. Trustore~\cite{trustore} proposes secure storage based on an FPGA as a faster alternative. While these solutions mitigate side-channels that reveal memory access patterns to the adversary in general, Invisimem~\cite{invisimem}, Obfusmem~\cite{obfusmem} only addresses the side-channels observable on the memory bus. 
Through \thename, we introduce another method for side-channel mitigation, in which the enclave user offload sensitive operations to a PIM to benefit from side-channel-free execution.

\subsection{Extending Trust to Peripheral Devices}
Existing works have presented peripheral device designs that create a secure communication channel with the host enclaves. Graviton~\cite{graviton} proposed a set of hardware modifications for GPU that enables secure acceleration for confidential computing in the cloud. HIX~\cite{hix}, sought to achieve the same goal but proposed modifications to the I/O interconnect between CPU and GPU. Trustore~\cite{trustore} uses an FPGA storage device to provide extra storage space that is free of memory side-channels for SGX. One design component these works have in common is the support for remote attestation and secure communication channels. Since achieving trusted I/O with a device is not viable in SGX's security model, the device itself must actively participate in establishing a secure channel with the processor enclave. Our design, \thename adapts many of the requirements for a device in extending processor enclave's trust to I/O devices generalized by the previous works.

\section{Conclusion}
We presented \thename, an architecture that enables confidential computing inside memory by retrofitting the processor-in-memory architecture. Through a careful security analysis of the new architecture, we proposed a set of non-intrusive yet imperative changes that guarantees the data's confidentiality and integrity computed inside PIM. We evaluated our design as well as a proof-of-concept application for our design using a cycle-accurate full system simulation. Our evaluation shows that the encrypted data transfer in our design only incurs only a $17.85\%$ overhead in the maximum throughput inside memory compared to the unmodified PIM architecture that does not support encryption. We also evaluated an secure $k$-mean computation program that runs on our architecture to accelerate data-intensive operations. We observed only 3.7\% increase in total execution time compared to offloading to unprotected PIM architecture.

\section*{Acknowledgment}
This work was supported by the National Research Foundation of Korea~(NRF) grant funded by the Korea government~(MSIT)~(NRF-2020R1C1C1011980).


%

\ifCLASSOPTIONcaptionsoff
  \newpage
\fi




\printbibliography
\balance
%
%




\end{document}